\documentclass[aip,jcp,reprint,onecolumn,amsmath,amssymb,amsfonts,preprintnumbers,showkeys,showpacs]{revtex4-1}
\usepackage{graphicx,dcolumn,bm,color,hyperref,multirow,microtype,amsmath,amssymb,rotating,cleveref}

\DeclareMathOperator{\Li}{Li}
\renewcommand{\bf}{\mathbf}
\newcommand{\eps}{\epsilon}
\newcommand{\EHF}{\epsilon_{\rm HF}}
\newcommand{\Ec}{\epsilon_{\rm c}}
\newcommand{\Eh}{E_{\rm h}}
\newcommand{\mEh}{{\rm m}E_{\rm h}}
\newcommand{\uEh}{{\mu}E_{\rm h}}
\newcommand{\mc}{\multicolumn}

\newcommand{\db}[2]{\langle #1 || #2 \rangle}
\newcommand{\alert}[1]{\textcolor{black}{#1}}

\begin{document}

\title{Uniform electron gases. I. Electrons on a ring}

\author{Pierre-Fran\c{c}ois Loos}
\email{loos@rsc.anu.edu.au}
\author{Peter M. W. Gill}
\email{peter.gill@anu.edu.au}
\affiliation{Research School of Chemistry, 
Australian National University, Canberra, ACT 0200, Australia}

\begin{abstract}
We introduce a new paradigm for one-dimensional uniform electron gases (UEGs).  In this model, $n$ electrons are confined to a ring and interact via a bare Coulomb operator.  We use Rayleigh-Schr\"odinger perturbation theory to show that, in the high-density regime, the ground-state reduced ({\it i.e.}~per electron) energy can be expanded as $\eps(r_s,n) = \eps_0(n) r_s^{-2} + \eps_1(n) r_s^{-1} + \eps_2(n) +\eps_3(n) r_s + \ldots$, where $r_s$ is the Seitz radius.  We use strong-coupling perturbation theory and show that, in the low-density regime, the reduced energy can be expanded as $\eps(r_s,n) = \eta_0(n) r_s^{-1} + \eta_1(n) r_s^{-3/2} + \eta_2(n) r_s^{-2} + \ldots$.  We report explicit expressions for $\eps_0(n)$, $\eps_1(n)$, $\eps_2(n)$, $\eps_3(n)$, $\eta_0(n)$ and $\eta_1(n)$ and derive the thermodynamic (large-$n$) limits of each of these.  Finally, we perform numerical studies of UEGs with $n = 2, 3, \ldots, 10$, using Hylleraas-type and quantum Monte Carlo methods, and combine these with the perturbative results to obtain a picture of the behavior of the new model over the full range of $n$ and $r_s$ values.
\end{abstract}


\maketitle

\section{Introduction}
In a recent paper, \cite{UEGs12} we showed that the traditional concept of the uniform electron gas (UEG), {\it i.e.}~a homogeneous system of finite density, consisting of an infinite number of electrons in an infinite volume, \cite{Vignale, 2DEG, 3DEG} is inadequate to model the UEGs that arise in finite systems. Accordingly, we proposed to embark on a comprehensive study of quasi-exact properties of finite-size UEGs, in order eventually to create improved approximations in density-functional theory. \cite{ParrBook}

In an earlier paper, \cite{Glomium11} we introduced an alternative paradigm, in which $n$ electrons are confined to a $D$-sphere (with $D \ge 2$), that is, the surface of a ($D+1$)-dimensional ball. These systems possess uniform densities, even for finite $n$ and, because all points on a $D$-sphere are equivalent, their mathematical analysis is relatively straightforward. \cite{TEOAS09, QuasiExact09, LoosConcentric, LoosHook, LoosExcitSph, Excitons12}  In the present paper, we study the one-dimensional ($D=1$) version of model, in which $n$ electrons are confined to a ring of radius $R$.  The electron density of this $n$-electron UEG, which we will call $n$-ringium, is  
\begin{equation}
	\rho = \frac{n}{2\pi R} = \frac{1}{2\,r_s},
\end{equation} 
where $r_s = \pi R/n$ is the Seitz radius. \alert{In this study, the high-density (small-$r_s$) limit is defined by $R\to0$ for fixed $n$, while the low-density (large-$r_s$) limit is defined by $R\to\infty$ for fixed $n$. We do not include a fictitious uniform positive background charge because, unlike the situation in 2D and 3D UEGs, its inclusion in 1D systems causes the Coulomb energy to diverge.}

In most previous work on the one-dimensional (1D) UEG, the true Coulomb potential $1/r_{12}$ has been avoided because of the intractability of its Fourier transform.  Instead, most workers have softened the potential, either by adding a transverse harmonic component \cite{Pederiva02, Casula06, Lee11a} or by using a potential of the form $1/\sqrt{r_{12}^2+\mu^2}$.  In the latter case, the $\mu$ parameter eliminates the singularity at $r_{12} = 0$ while retaining the long-range Coulomb tail. \cite{Schulz93, Fogler05a, Lee11a}

However, the introduction of a parameter $\mu > 0$ is undesirable, for it modifies the physics of the system in the high-density regime where neighboring electrons repel far too weakly.  It is also unnecessary, because the true Coulomb potential is so repulsive that it causes the wave function to vanish when any two electrons touch, thereby removing the possibility of an energy divergence. \cite{Astrakharchik11} \alert{For 1D systems, we have recently shown that the exact wave function $\Psi$ behaves as
\begin{equation}
	\Psi(r_{12}) = r_{12} \left(1+ \frac{r_{12}}{2}\right) + O(r_{12}^3)
\end{equation}
for small $r_{12}$, \cite{QR12} which is the 1D analog of the (three-dimensional) Kato cusp condition. \cite{Kato57}}

This nodal behavior leads to the 1D Bose-Fermi mapping \cite{Girardeau60} which states that the ground state wave function of the bosonic (B) and fermionic (F) states are related by $\Psi_\text{B} (\bf{R}) = \left| \Psi_\text{F} (\bf{R})\right|$, where $\bf{R} = (\bf{r}_1,\bf{r}_2,\ldots,\bf{r}_n)$ are the one-particle coordinates. In case of bosons, the divergence of the Coulomb potential has the effect of mimicking the Pauli principle which prohibits two fermions from touching.  This implies that, for 1D systems, the bosonic and fermionic ground states are degenerate and the system is ``spin-blind''. Consequently, the paramagnetic and ferromagnetic states are degenerate and we will consider only the latter. \cite{Lee11a}

The electrons-on-a-ring paradigm has been intensively studied as a model for quantum rings (QRs), which are tiny, self-organised, ring-shaped semiconductors \cite{Warburton00, Fuhrer01} characterised by three parameters: radius ($R$), width ($\delta$) and electron number ($n$). Modern microfabrication technology has yielded InGaAs and GaAlAs/GaAs QRs that bind only a few electrons, \cite{Lorke00, Keyser03} in contrast with the mesoscopic rings on GaAs which hold much larger numbers of electrons. \cite{Mailly93} These low-dimensional systems are the subject of considerable scientific interest and have been intensively studied, both experimentally \cite{Mailly93, Warburton00, Lorke00, Fuhrer01, Bayer03, Keyser03, Fuhrer04, Sigrist04} and theoretically, \cite{Viefers04, Fogler05b, Fogler06, Niemela96, Gylfadottir06, Pederiva02, Emperador03, Emperador01, Rasanen09, Aichinger06, Manninen09, QR12} mainly because of the observation of  Aharonov-Bohm oscillations. \cite{Aharonov59, Aronov93, Morpurgo98, Emperador03}

As a first approximation, QRs can be modelled by electrons confined to a perfect ring (\textit{i.e.}~$\delta=0$).  In a recent paper, \cite{QR12} we considered a pair of electrons (\textit{i.e.}~$n=2$) on such a ring and discovered that their Schr\"odinger equation can be solved exactly, provided that the radius takes one of an infinite number of special values.  Some of the solutions exhibit the Berry phase phenomenon, \textit{i.e.}~if one of the electrons moves once around the ring and returns to its starting point, the wave function of the system changes sign. QRs are among the simplest systems with this peculiar property. 

In Section \ref{sec:Pert}, we first use Rayleigh-Schr\"odinger perturbation theory to investigate the energy in the high-density regime \cite{GellMann57} and then strong-coupling perturbation theory to study the low-density regime, where the electrons form a Wigner crystal. \cite{Wigner34}  In Section \ref{sec:EC}, we use explicitly correlated (EC) methods to determine the energy of $n$-ringium for $n = 2, 3, 4, 5$.  These methods are accurate for very small $n$ but their cost grows very rapidly with $n$.  In Section \ref{sec:QMC}, we turn to quantum Monte Carlo (QMC) approaches for studying $n$-ringium up to $n=10$  over a range of densities. These methods provide a different approach to the many-body problem:  variational Monte Carlo (VMC) \cite{McMillan65, Ceperley77, Umrigar99} and diffusion Monte Carlo (DMC) \cite{Kalos74, Ceperley79, Reynolds82} methods can be used to treat systems in one and higher dimensions at a computational cost that grows relatively slowly with $n$ (at least, when $n$ is not too large. \cite{Nemec10})

We frame our discussion in terms of \emph{reduced} energy $\eps(r_s,n)$, {\it i.e.}~energy per electron, so that we can pass smoothly from finite to infinite $n$.  One of the key goals of the paper is to develop an understanding of the correlation energy, which is defined as the difference
\begin{equation}
	\Ec(r_s,n) = \eps(r_s,n) - \EHF(r_s,n), 
\end{equation}
between the exact and Hartree-Fock (HF) energies.  Atomic units are used throughout, but we report total energies in hartrees ($\Eh$) and correlation energies in millihartrees ($\mEh$).

\section{Perturbative methods} \label{sec:Pert} 
\subsection{High-density expansion} 
The Hamiltonian of the system is
\begin{equation} \label{eq:H}
	H = - \frac{1}{2R^2} \sum_{i=1}^{n} \frac{\partial^2}{\partial\theta_i^2} + \sum_{i<j}^{n} \frac{1}{r_{ij}},
\end{equation}
where $\theta_i$ is the angle of electron $i$ around the ring center, and
\begin{equation}
	r_{ij} = \left| \bf{r}_i - \bf{r}_j \right| = R \sqrt{2-2\cos(\theta_i - \theta_j)}
\end{equation}
is the across-the-ring distance between electrons $i$ and $j$.

In the high-density ({\it i.e.}~small $r_s$) regime, the kinetic energy is dominant and it is natural to define a zeroth-order Hamiltonian 
\begin{equation}
	H_0 = - \frac{1}{2R^2} \sum_{i=1}^{n} \frac{\partial^2}{\partial\theta_i^2},
\end{equation}
and a perturbation
\begin{equation}
	V = \sum_{i<j}^{n} r_{ij}^{-1}.
\end{equation}
The non-interacting orbitals and orbital energies are
\begin{gather}
	\chi_a(\theta) = (2\pi R)^{-1/2} \exp(i\,a\,\theta),		\\
	\kappa_a = \frac{a^2}{2R^2},
\end{gather}
where
\begin{equation}
	a = \begin{cases}
			\ldots, -2, -1, 0, +1, +2, \ldots,									&	\text{if $n$ is odd,}	\\
			\ldots, -\frac{3}{2}, -\frac{1}{2}, +\frac{1}{2}, +\frac{3}{2}, \ldots,	&	\text{if $n$ is even.}
		\end{cases}
\end{equation}
A Slater determinant $^n\Psi_i$ of any $n$ of these orbitals has an energy $E_i$ and is an antisymmetric eigenfunction of $H_0$. In the lowest energy (aufbau) determinant $^n\Psi_0$, we occupy the orbitals with
\begin{equation} \label{eq:occ}
	a = -\frac{n-1}{2}, -\frac{n-3}{2}, \ldots, +\frac{n-3}{2}, +\frac{n-1}{2}.
\end{equation}
Following the approach of Mitas, \cite{Mitas06} one discovers the remarkable result 
\begin{equation} \label{eq:PsiHF}
	^n\Psi_0 \propto \prod_{i<j}^n \Hat{r}_{ij},
\end{equation}
where 
\begin{equation}
	\Hat{r}_{ij} = 2 R \sin\left(\frac{\theta_i-\theta_j}{2}\right)
\end{equation} 
is a signed interelectronic distance.  It follows immediately that $^n\Psi_0$ has a node whenever $\theta_i = \theta_j$ and, therefore, possesses the same nodes as the exact wave function.  This will have important ramifications in Section \ref{sec:QMC}.

Rayleigh-Schr\"odinger theory yields the perturbation expansion for the reduced energy
\begin{equation} \label{eq:expansion}
	\eps(r_s,n) = \frac{\eps_0(n)}{r_s^2} + \frac{\eps_1(n)}{r_s} + \eps_2(n) + \eps_3(n) r_s + \ldots,
\end{equation}
where the high-density coefficients $\eps_j(n)$ are found by setting $R=1$ and evaluating
\begin{subequations}
\begin{align}
	\eps_0(n) & = \frac{\pi^2}{n^3} \langle \Psi_0 | H_0 | \Psi_0 \rangle,														\label{eq:eps0}		\\
	\eps_1(n) & = \frac{\pi}{n^2} \langle \Psi_0 | V | \Psi_0 \rangle,															\label{eq:eps1}		\\
	\eps_2(n) & = \frac{1}{n} \sum_i \frac{ \langle \Psi_0 | V | \Psi_i \rangle \langle \Psi_i | V | \Psi_0 \rangle}{E_0 - E_i},	\label{eq:eps2}		\\
	\eps_3(n) & = \frac{1}{\pi} \sum_i \sum_j \frac{ \langle \Psi_0 | V | \Psi_i \rangle \langle \Psi_i | V-n^2\eps_1/\pi | \Psi_j \rangle \langle \Psi_j | V | \Psi_0 \rangle}
																			{(E_0 - E_i)(E_0 - E_j)}.						\label{eq:eps3}
\end{align}
\end{subequations}

\subsubsection{Double-bar integrals}
To evaluate the coefficients $\eps_j(n)$ with $j > 0$, one requires the ``double-bar'' integrals
\begin{equation}
	\db{ab}{cd} = \int_0^{2\pi} \!\!\! \int_0^{2\pi} \frac{\chi_a^*(\theta_1) \chi_b^*(\theta_2) [ \chi_c(\theta_1)\chi_d(\theta_2) - \chi_c(\theta_2)\chi_d(\theta_1)]}
															{r_{12}} d\theta_1 d\theta_2.
\end{equation}
By elementary integration, one can show that
\begin{equation}
	\db{ab}{cd} = \begin{cases}
						V_{c-b,c-a},		&	a+b = c+d,	\\
						\qquad 0,		&	\text{otherwise,}
					\end{cases}
\end{equation}
where
\begin{equation}
	V_{p,q} = \frac{1}{\pi} \left[ \psi(p+\tfrac{1}{2}) - \psi(q+\tfrac{1}{2}) \right],
\end{equation}
and $\psi$ is the digamma function. \cite{NISTbook}

\begin{table}
	\caption{\label{tab:eps} High-density coefficients for $n$-ringium.  ($\zeta(3)$ is Ap\'ery's constant. \cite{NISTbook})}
	\begin{ruledtabular}
	\begin{tabular}{ccccc}
		\noalign{\smallskip}
		$n$	&	$\eps_0(n)$			&	$\eps_1(n)$					&			$\eps_2(n)$			&		$\eps_3(n)$				\\
		\noalign{\smallskip} \hline \noalign{\smallskip}
		2	&	$\frac{1}{32}\pi^2$		&	$\frac{1}{2}$
										&	$1 - \frac{10}{\pi^2}$
										&	$\frac{8(12 \ln 2 - 19)}{3\pi^2} + \frac{16(26 - 7\zeta(3))}{\pi^4}$										\\[2mm]
		3	&	$\frac{1}{27}\pi^2$		&	$\frac{20}{27}$
										&	$\frac{16}{9} - \frac{1436}{81\pi^2}$
										&	$\frac{8(1080 \ln 2 - 997)}{81\pi^2} + \frac{8(13046 - 4725\zeta(3))}{243\pi^4}$																							\\[2mm]
		4	&	$\frac{5}{128}\pi^2$	&	$\frac{9}{10}$
										&	$\frac{109}{45} - \frac{244168}{10125\pi^2}$
										&	0.00487354																							\\[2mm]
		5	&	$\frac{1}{25}\pi^2$		&	$\frac{892}{875}$
										&	$\frac{4688}{1575} - \frac{514012364}{17364375\pi^2}$
										&	0.00556461																							\\[2mm]
		6	&	$\frac{35}{864}\pi^2$	&	$\frac{6323}{5670}$
										&	$\frac{2339}{675} - \frac{461265158}{13395375\pi^2}$
										&	0.00605813																							\\[2mm]
		7	&	$\frac{2}{49}\pi^2$		&	$\frac{13528}{11319}$
										&	$\frac{1420256}{363825} - \frac{33870168846728}{873632962125\pi^2}$
										&	0.00642454																							\\[2mm]
		8	&	$\frac{21}{512}\pi^2$	&	$\frac{7591}{6006}$
										&	$\frac{20349053}{4729725} - \frac{81975019672689056}{1919371617788625\pi^2}$
										&	0.00670533																							\\[2mm]
		9	&	$\frac{10}{243}\pi^2$	&	$\frac{4831544}{3648645}$
										&	$\frac{66244064}{14189175} - \frac{266761139809046216}{5758114853365875\pi^2}$
										&	0.00692616																							\\[2mm]
		10	&	$\frac{33}{800}\pi^2$	&	$\frac{2512297}{1823250}$
										&	$\frac{1207979879}{241215975} - \frac{7026989855398034506022}{141448091372932719375\pi^2}$
										&	0.00710359																							\\[2mm]
		\hline \noalign{\smallskip}
		$\infty$	&	$\pi^2/24$			&	$\ln \sqrt n$
										&	$-\pi^2/360$
										&	$0.00844621$																							\\[2mm]
	\end{tabular}
	\end{ruledtabular}
\end{table}

\subsubsection{Zeroth order}
The zeroth-order coefficient \eqref{eq:eps0} becomes
\begin{equation}
	\eps_0(n) = \frac{\pi^2}{n^3} \sum_a^\text{occ} \frac{a^2}{2},
\end{equation}
where the ``occ'' indicates sums over all occupied orbitals \eqref{eq:occ}, and this reduces to
\begin{equation}
	\eps_0(n) = \frac{n^2-1}{n^2} \frac{\pi^2}{24}.
\end{equation}
In the thermodynamic ({\it i.e.}~$n\to\infty$) limit, this approaches
\begin{equation}
	\eps_0 = \frac{\pi^2}{24},
\end{equation}
which is identical to the kinetic energy coefficient in the ideal Fermi gas in 1D. \cite{Vignale, Glomium11}

\subsubsection{First order}
The first-order coefficient \eqref{eq:eps1} becomes
\begin{equation}
	\eps_1(n) = \frac{\pi}{n^2} \sum_{a<b}^\text{occ} \db{ab}{ab},
\end{equation}
which can be reduced to
\begin{equation}
	\eps_1(n) = \left(\frac{1}{2} - \frac{1}{8n^2}\right) \left[ \psi(n+\tfrac{1}{2}) - \psi(\tfrac{1}{2}) \right] - \frac{3}{4}.
\end{equation}
This can be found in closed form for any $n$ (see Table \ref{tab:eps}).  Because of the slow decay of the Coulomb operator, the coefficient grows logarithmically with $n$ and it can be shown that
\begin{equation}
\label{eps1}
	\eps_1(n) \sim \ln\sqrt{n} + (\ln 2 + \gamma/2 - 3/4) + O(n^{-2}\ln n),
\end{equation}
where $\gamma$ is the Euler-Mascheroni constant. \cite{NISTbook}

The sum of the first two terms in \eqref{eq:expansion} gives the HF energy of $n$-ringium
\begin{equation}
	\eps_\text{HF}(r_s,n) = \frac{\eps_0(n)}{r_s^2} + \frac{\eps_1(n)}{r_s}.
\end{equation}

\subsubsection{Second order}
The second-order coefficient \eqref{eq:eps2} becomes
\begin{equation} \label{eq:MP2}
	\eps_2(n) = - \frac{1}{n} \sum_{a<b}^{\text{occ}}\sum_{r<s}^{\text{virt}} \frac{ \db{ab}{rs} \db{rs}{ab}}{\kappa_r + \kappa_s - \kappa_a - \kappa_b},
\end{equation}
where the ``virt'' indicates sums over all virtual orbitals.  If the double-bar integrals do not vanish, {\it i.e.}~$a+b = r+s$, then 
\begin{equation}
	\kappa_r + \kappa_s - \kappa_a - \kappa_b = (r-a)(r-b),
\end{equation}
and we obtain
\begin{equation} \label{eq:MP2-final}
	\eps_2(n) = - \frac{1}{n} \sum_{a<b}^\text{occ} \sum_{r=r_\text{min}}^\infty \frac{V_{r-a,r-b}^2}{(r-a)(r-b)},
\end{equation}
where
\begin{equation}
	r_\text{min} = \frac{n+1}{2}+\max(a+b,0).
\end{equation}
The sums in \eqref{eq:MP2-final} can be evaluated in closed form for any $n$ (see Table \ref{tab:eps}).  In the $r_s \to 0$ limit, the higher terms in \eqref{eq:expansion} vanish and the $\eps_2(n)$ expressions in Table \ref{tab:eps} are therefore the \emph{exact} correlation energies of infinitely dense $n$-ringium.

In the thermodynamic limit, $\eps_2(n)$ approaches
\begin{align}
	\eps_2 	& = - \lim_{n\to\infty} \frac{1}{n} \sum_{a<b}^\text{occ} \sum_{r=r_\text{min}}^\infty
				\frac{\left[\frac{1}{\pi} \ln\left( \frac{r-a}{r-b}\right)\right]^2}{(r-a)(r-b)}							\notag	\\
			& = - \frac{1}{3\pi^2} \int_0^1 \!\! \int_{-x}^x \frac{1}{x-y} \ln^3 \left(\frac{1+x}{1+y}\right) \,dx \,dy	\notag	\\
			& = - \frac{\pi^2}{360},
\end{align}
which implies that, in the dual thermodynamic/high-density limit, the exact correlation energy of ringium is $-27.4\ \mEh$ per electron.  The same value of $\eps_2$  can be derived for 1D jellium, \cite{1DEG} affirming the equivalence of the electrons-on-a-ring and electrons-on-a-wire models in the thermodynamic limit. \cite{Glomium11}

Using a quasi-1D model with a transverse harmonic potential, Casula {\it et al.}~were led to conclude that, in the same limit, the correlation energy vanishes. \cite{Casula06}  This qualitatively different prediction stresses the importance of employing a realistic Coulomb operator for high-density UEGs.

\subsubsection{Third order}
The third-order coefficient \eqref{eq:eps3} becomes
\begin{align} \label{eq:eps3n}
	\eps_3(n)	& = \frac{1}{8\pi} \sum_{abcd}^\text{occ} \sum_{rs}^\text{virt} \frac{\db{ab}{rs} \db{rs}{cd} \db{cd}{ab}}{(r-a)(r-b)(r-c)(r-d)}
				    + \frac{1}{8\pi} \sum_{ab}^\text{occ} \sum_{rstu}^\text{virt} \frac{\db{ab}{rs} \db{rs}{tu} \db{tu}{ab}}{(r-a)(r-b)(t-a)(t-b)}		\notag	\\
				& + \frac{1}{\pi} \sum_{abc} ^\text{occ} \sum_{rst}^\text{virt} \frac{\db{ab}{rs} \db{cs}{tb} \db{rt}{ac}}{(r-a)(r-b)(r-a)(r-c)}
				    + \frac{1}{\pi} \sum_{abc}^\text{occ} \sum_{rst}^\text{virt} \frac{\db{ab}{rs} \db{ar}{ct} \db{rs}{ab}}{(r-a)(r-b)(r-a)(r-b)},
\end{align}
and, like $\eps_2(n)$, this can be rewritten in terms of products of $V_{p,q}$.  The expression is cumbersome but can be evaluated in closed form for any $n$ and Table \ref{tab:eps} illustrates this for $n = 2$ and 3.

In the thermodynamic limit, $\eps_3(n)$ approaches the numerical value
\begin{equation} \label{eq:eps3inf}
	\eps_3 = +0.00844621,
\end{equation}
but we have been unable to obtain this in closed form.  Numerical evidence suggests \cite{1DEG} that \eqref{eq:eps3inf} is also true of 1D jellium.

\alert{Interestingly, second- and third-order perturbation theories applied to 1D jellium do not encounter divergence issues as in 2D and 3D jellium, where one has to use resummation techniques to produce finite results. \cite{Rajagopal77, GellMann57} The divergence occurs from third order and second order for 2D jellium and 3D jellium, respectively. In the case of 1D jellium, every terms of the perturbation expansion seem to converge.}

\subsection{Low-density expansion} 
In the low-density ($r_s \gtrsim 2$) regime, \cite{Gylfadottir06} the electrons form a Wigner crystal. Using strong-coupling perturbation theory, \cite{TEOAS09} the energy can be written
\begin{equation}
\label{eq:large-rs}
	\epsilon(r_s,n) = \frac{\eta_0(n)}{r_s} + \frac{\eta_1(n)}{r_s^{3/2}} + \ldots,
\end{equation}
where the first term represents the classical Coulomb energy of the static electrons and the second is their harmonic zero-point vibrational energy.

The Wigner crystal, which is the solution to the 1D Thomson problem, \cite{Thomson04} consists of $n$ electrons separated by an angle $2\pi/n$ and yields
\begin{equation} \label{eq:EW}
	\eta_0(n) = \frac{\pi}{2n^2} \sum_{k=1}^{n-1} \frac{n-k}{\sin(k\pi/n)}.
\end{equation}
The second term in the expansion \eqref{eq:large-rs} is found by summing the frequencies of the normal modes obtained by diagonalization of the Hessian matrix. For electrons on a ring, the Hessian is circulant and its eigenvalues and eigenvectors can be found in compact form, yielding 
\begin{equation} 
\label{eq:eta32}
	\eta_1(n) =\frac{\pi^{3/2}}{4n^{5/2}} \sum_{i=1}^{n-1} \sqrt{\sum_{k=1}^{n-1} \frac{2-\sin^2(k\pi/n)}{\sin^3(k\pi/n)} \sin^2(ik\pi/n)}.
\end{equation}	

In the thermodynamic limit, one finds that
\begin{equation}
	\eta_0 = \ln \sqrt{n} + \frac{\ln (2/\pi)+\gamma}{2} + o(n^0),
\end{equation}
which has the same logarithmic divergence as $\eps_1$, but with a different constant term.  Likewise, one can show that
\begin{align} 
	\eta_1	& = \frac{\pi^{3/2}}{4n^{5/2}} \sum_{i=1}^{n-1} \sqrt{\sum_{k=1}^{\infty} \frac{4}{(k\pi/n)^3} \sin^2(ik\pi/n)}	\notag	\\
			& = \frac{1}{4\pi} \int_0^{\pi} \sqrt{2\Li_3(1) - \Li_3(e^{i \theta}) - \Li_3(e^{-i \theta})}\,d\theta,
\end{align}	
where $\Li_3$ is the trilogarithm function. \cite{NISTbook} We have not been able to find this integral in closed form, but it can be computed numerically with high precision, and yields $\eta_1 = 0.359933$, which is identical to the value found by Fogler \cite{Fogler05a} for an infinite ultrathin wire and a potential of the form $1/\sqrt{r_{12}^2+\mu^2}$. This shows that, unlike the high-density limit where the details of the interelectronic potential are critically important, the correct low-density result can be obtained by using a softened Coulomb potential.

Thus, in the dual thermodynamic/low-density region, we have
\begin{equation} \label{eq:Ec_lo}
	\Ec(r_s) =  -\frac{\ln(\sqrt{2\pi})-3/4}{r_s} + \frac{0.359933}{r_s^{3/2}} + O(r_s^{-2}).
\end{equation}
The same expansion can be derived for the infinite wire, \cite{Fogler05a} confirming the equivalence of the electrons-on-a-ring and electrons-on-a-wire models in the thermodynamic limit. \cite{Glomium11}

\section{Explicitly correlated methods} \label{sec:EC}
Because the full set of interelectronic distances $r_{ij}$ determine the positions of the electrons to within an overall rotation that is irrelevant in the ground state, it is appropriate to adopt these variables as natural coordinates and to expand the correlated wave function in terms of these distances.

\subsection{2-ringium}
\begin{table}
	\caption{\label{tab:2e} Convergence with $M$ of the energy of 2-ringium with $r_s = 1$.}
	\begin{ruledtabular}
	\begin{tabular}{ccc}
		$M$	&		$\eps(1,2)$				&		$-\Ec(1,2)$			\\
		\hline
		0		&	0.808\,425\,137\,534		&	0						\\
		1		&	0.797\,201\,143\,955		&	11.223\,993\,579		\\
		2		&	0.797\,175\,502\,306		&	11.249\,635\,229 		\\
		3		&	0.797\,175\,223\,852		&	11.249\,913\,682		\\
		4		&	0.797\,175\,219\,345		&	11.249\,918\,190		\\
		5		&	0.797\,175\,219\,257		&	11.249\,918\,277		\\	
		6		&	0.797\,175\,219\,255		&	11.249\,918\,279		\\
	\end{tabular}
	\end{ruledtabular}
\end{table}

The HF wave function for 2-ringium is
\begin{equation}
	^2\Psi_0 = \Hat{r}_{12}.
\end{equation}
In the light of its simplicity, and following our previous analysis of the quasi-exact solutions, \cite{QR12} it is natural to consider correlated wave functions that are products of $^2\Psi_0$ and a correlation factor, {\it viz.}
\begin{equation} \label{eq:Psi2e}
	^2\Psi_M = {^2\Psi_0} \sum_{m=0}^{M} c_m r_{12}^m.
\end{equation}
The overlap, kinetic and potential matrix elements can be found as outlined in the Appendix.

Table \ref{tab:2e} shows the energies obtained by solving the secular eigenvalue problem for $r_s = 1$.  They converge rapidly, with $M$ = 1, 2, 4, 6 yielding milli-, micro-, nano- and pico-hartree accuracy, respectively.  It is interesting to compare the correlation energy ($-11\ \mEh$) with the corresponding value ($-114\ \mEh$) for two electrons on a 2D sphere. \cite{Frontiers10}  One normally expects the correlation energy to decrease in higher dimensions \cite{EcProof10} but the 1D case is anomalous because the HF wavefunction \eqref{eq:PsiHF} places the two electrons in different orbitals.

The key discovery from this investigation is that including just the linear ($r_{12}$) and quadratic ($r_{12}^2$) terms in the expansion \eqref{eq:Psi2e} affords microhartree accuracy for the energy of 2-ringium.  We now ask whether this is true for larger values of $n$. 

\subsection{3-ringium}
The HF wave function for 3-ringium is
\begin{equation}
	^3\Psi_0 = \Hat{r}_{12}\,\Hat{r}_{13}\,\Hat{r}_{23},
\end{equation}
and we have explored both Hylleraas-type wavefunctions \cite{Hylleraas29, Hylleraas30, Hylleraas64}
\begin{subequations}
\begin{gather}
	^3\Psi_M^\text{Hy} = {^3\Psi_0} \sum_{m=0}^M \ \ \sum_{i+2j+3k \le m} c_{ijk}\,s_1^i\,s_2^j\,s_3^k,	\\
	s_1 = r_{12} + r_{13} + r_{23},																	\\
	s_2 = r_{12}\,r_{13} + r_{12}\,r_{23} + r_{13}\,r_{23},												\\
	s_3 = r_{12}\,r_{13}\,r_{23},
\end{gather}
\end{subequations}
and Jastrow-type wavefunctions \cite{Jastrow55}
\begin{equation}
	^3\Psi_M^\text{Ja} = {^3\Psi_0} \left( \sum_{m=0}^M c_m r_{12}^m \right) \left( \sum_{m=0}^M c_m r_{13}^m \right) \left( \sum_{m=0}^M c_m r_{23}^m \right).
\end{equation}
The required matrix elements can be found as outlined in the Appendix.

\begin{table}
	\caption{\label{tab:3e-M} Convergence with $M$ of the energy of 3-ringium with $r_s = 1$.}
	\begin{ruledtabular}
	\begin{tabular}{ccccc}	
				&			\mc{2}{c}{Hylleraas expansion}					&			\mc{2}{c}{Jastrow expansion}				\\
				\cline{2-3}\cline{4-5}
		$M$	&		$\eps(1,3)$			&		$-\Ec(1,3)$				&		$\eps(1,3)$			&		$-\Ec(1,3)$			\\
		\hline				
		0		&	1.106\,281\,644\,485	&	0							&	1.106\,281\,644\,485	&	0						\\
		1		&	1.091\,649\,204\,702	&	14.632\,439\,783			&	1.090\,999\,267\,912	&	15.282 376 573		\\
		2		&	1.090\,936\,176\,037	&	15.345\,468\,448			&	1.090\,936\,808\,374	&	15.344\,836\,111		\\
		3		&	1.090\,935\,619\,110	&	15.346\,025\,375			&	1.090\,936\,772\,712	&	15.344\,871\,773		\\
		4		&	1.090\,935\,608\,007	&	15.346\,036\,478			&	1.090\,936\,607\,858	&	15.345\,036\,627		\\
		5		&	1.090\,935\,607\,817	&	15.346\,036\,667			&	1.090\,936\,593\,657	&	15.345\,050\,828		\\
		6		&	1.090\,935\,607\,811	&	15.346\,036\,674			&	1.090\,936\,589\,183	&	15.345\,055\,301		\\
		7		&	1.090\,935\,607\,810	&	15.346\,036\,674			&	1.090\,936\,588\,261	&	15.345\,056\,224		\\
	\end{tabular}
	\end{ruledtabular}
\end{table}

The Hylleraas expansion converges rapidly for 3-ringium with $r_s = 1$ and Table \ref{tab:3e-M} reveals that, as in 2-ringium, $M$ = 1, 2, 4, 6 yields milli-, micro-, nano- and pico-hartree accuracies, respectively.  The reduced correlation energy is roughly 35\% greater than that in 2-ringium.  Because of its factorized form, the limiting Jastrow energy is $\approx 1\ \uEh$ above the exact value.

\subsection{4- and 5-ringium}
The HF wave functions for 4- and 5-ringium, respectively, are
\begin{gather}
	^4\Psi_0 = \Hat{r}_{12}\,\Hat{r}_{13}\,\Hat{r}_{14}\,\Hat{r}_{23}\,\Hat{r}_{24}\,\Hat{r}_{34},		\\
	^5\Psi_0 = \Hat{r}_{12}\,\Hat{r}_{13}\,\Hat{r}_{14}\,\Hat{r}_{15}\,\Hat{r}_{23}\,\Hat{r}_{24}\,\Hat{r}_{25}\,\Hat{r}_{34}\,\Hat{r}_{35}\,\Hat{r}_{45}.
\end{gather}
Hylleraas calculations on these systems are complicated because of the large number of many-electron integrals which are required.  Nonetheless, we were able to perform such calculations, up to $M = 2$ for 4-ringium and up to $M = 1$ for 5-ringium, and the results are summarized in Table \ref{tab:4e-5e-M}.  It is important to allow $r_{ij} r_{kl}$ terms (which couple two electron pairs) and $r_{ij} r_{ik}$ terms (which describe three-electron interactions) to have distinct Hylleraas coefficients:  failing to do so raises the energy by $\approx 1\ \uEh$.  The energies in Table \ref{tab:4e-5e-M} are higher than our best estimates (see Table \ref{tab:rs1}) by roughly 2 $\mEh$ (for $M = 1$) and 50 $\uEh$ (for $M = 2$).

\begin{table}
	\caption{\label{tab:4e-5e-M} Convergence with $M$ of the energies of 4- and 5-ringium with $r_s = 1$.}
		\begin{ruledtabular}
		\begin{tabular}{ccccc}
					&			\mc{2}{c}{4-ringium}		&		\mc{2}{c}{5-ringium}			\\
					\cline{2-3}\cline{4-5}
		$M$	&	$\eps(1,4)$		&	$-\Ec(1,4)$		&	$\eps(1,5)$		&	$-\Ec(1,5)$		\\
		\hline
		0		&	1.285\,531		&	0				&	1.414\,213		&	0				\\	
		1		&	1.269\,785		&	15.746			&	1.398\,192 		&	16.021			\\
		2		&	1.268\,259		&	17.272			&	---				&	---				\\	
	\end{tabular}
	\end{ruledtabular}
\end{table}

\section{Quantum Monte Carlo methods} \label{sec:QMC}
\subsection{Variational Monte Carlo}
In the VMC method, the expectation value of the Hamiltonian with respect to a trial wave function is obtained using a stochastic integration technique. Within this approach a variational trial wave function $\Psi_\text{T}(\bf{R},\bf{c})$ is introduced, where $\bf{c} = (c_1,c_2,\ldots,c_M)$ are variational parameters.  One then minimizes the energy
\begin{equation}
	\eps_\text{VMC} = \frac{1}{n} \frac{\int \Psi_\text{T}(\bf{R},\bf{c}) H \Psi_\text{T}(\bf{R},\bf{c}) d\bf{R}}{\int \Psi_\text{T}(\bf{R},\bf{c})^2 d\bf{R}},
\end{equation}
with respect to the parameters $\bf{c}$ using the Metropolis Monte Carlo method of integration. \cite{Umrigar99} The resulting VMC energy is an upper bound to the exact ground-state energy, within the Monte Carlo error.  Unfortunately, any resulting observables are biased by the form of the trial wave function, and the method is therefore only as good as the chosen $\Psi_\text{T}$.

Here, we use electron-by-electron sampling with a transition probability density given by a Gaussian centered on the initial electron position. The VMC time step, which is the variance of the transition probability, is chosen to achieve a 50\% acceptance ratio. \cite{Lee11b}

\subsection{Diffusion Monte Carlo}
DMC is a stochastic projector technique for solving the many-body Schr\"odinger equation. \cite{Kalos74, Ceperley79, Reynolds82}  Its starting point is the time-dependent Schr\"odinger equation in imaginary time
\begin{equation} \label{eq:DMC}
	\frac{\partial \Psi(\bf{R},\tau)}{\partial \tau} = (H - S) \Psi(\bf{R},\tau),
\end{equation}
and it is exact, within statistical errors.  For $\tau \to \infty$, the steady-state solution of Eq.~\eqref{eq:DMC} for $S$ close to the ground-state energy is the ground-state $\Psi(\bf{R})$. \cite{Kolorenc11}  DMC generates configurations distributed according to the product of the trial and exact ground-state wave functions.  If the trial wave function has the correct nodes, the DMC method yields the exact energy, within a statistical error that can be made arbitrarily small by increasing the number of Monte Carlo steps. Thus, as in VMC, a high quality trial wave function is essential in order to achieve high accuracy. \cite{Umrigar93, Huang97}

Our DMC code follows the implementation of Reynolds {\it et al.}, \cite{Reynolds82} using a population of $\sim 5000$ walkers for each calculation. We have carefully checked that the population-control bias is negligible. The dependence of the energy upon the DMC time step $\Delta\tau$ was also investigated and the extrapolated value of the energy at $\Delta\tau = 0$ is obtained by a linear extrapolation. The number of points used in the fitting procedure depends on $r_s$. A minimum of 4 points has been used for linear interpolation in the set $\Delta\tau=$ 0.0001, 0.0002, 0.0005, 0.001, 0.002 and 0.005.  The extrapolated standard error is obtained by assuming that the data follow a Gaussian distribution. \cite{Lee11b} We note that the algorithm developed in Ref.~\onlinecite{Umrigar93} does not significantly reduce the time-step error in the present case. 

\subsection{Trial wave functions}
We have employed Jastrow trial wave functions
\begin{equation} \label{eq:PsiT}
	^n\Psi_M = {^n\Psi_0} \prod_{i<j}^n  \sum_{k=1}^{M} c_k r_{ij}^k,
\end{equation}
choosing $M = 5$ in order to obtain microhartree energy accuracy for $r_s = 1$.  The coefficients $c_k$ were optimized using Newton's method following the methodology developed by Umrigar and co-workers. \cite{Umrigar05, Toulouse07}  For $r_s \le 1$, we used energy minimization;  for $r_s > 1$, energy minimization was unstable and we minimized the variance of the local energy. \cite{Umrigar05}

\subsection{Fixed-node approximation}
DMC algorithms can be frustrated by the sign problem in fermionic systems. \cite{Loh90, Troyer05, Umrigar07}  To avoid this, it is common to apply the fixed-node approximation, \textit{i.e} to write the wave function as the product of a non-negative function and a function with a fixed nodal surface. \cite{Ceperley91} The DMC method then finds the best energy \emph{for that chosen nodal surface}, providing an upper bound for the ground-state energy. The exact ground-state energy is reached only if the nodal surface is exact but, fortunately for us, the nodal surface of the HF wave function Eq.~\eqref{eq:PsiHF} is exact and, therefore, DMC calculations using the trial wave function \eqref{eq:PsiT} yield the exact energy.  We have no fixed-node error.

\subsection{Results and discussion}
Table \ref{tab:rs1} summarizes the results of a systematic study of $n$-ringium systems with $r_s = 1$.  In all cases, our DMC calculations yielded energies with statistical uncertainties within 1 $\uEh$ and this allowed us to assess the accuracies of our explicitly correlated calculations.

\begin{table}
	\caption{\label{tab:rs1} Hartree-Fock, Explicitly Correlated and Diffusion Monte Carlo energies of $n$-ringium with $r_s = 1$.  Statistical errors in the last digit of the DMC energies are shown in parentheses.}
	\begin{ruledtabular}
	\begin{tabular}{cccccc}	
		$n$		&	$\EHF(1,n)$	&	$\eps_\text{EC}(1,n)$	&	$\eps_\text{DMC}(1,n)$	\\
		\hline
		2		&	0.808\,425		&		0.797\,175			&		0.797\,175(0)			\\
		3		&	1.106\,282		&		1.090\,936			&		1.090\,936(1)			\\
		4		&	1.285\,531		&		1.268\,259			&		1.268\,212(1)			\\
		5		&	1.414\,213		&		1.398\,192			&		1.395\,774(1)			\\
		6		&	1.514\,978		&			---				&		1.495\,841(1)			\\
		7		&	1.598\,000		&			---				&		1.578\,393(1)			\\
		8		&	1.668\,711		&			---				&		1.648\,770(1)			\\
		9		&	1.730\,359		&			---				&		1.710\,172(1)			\\
		10		&	1.785\,044		&			---				&		1.764\,671(1)			\\
	\end{tabular}
	\end{ruledtabular}
\end{table}

\begin{figure}
\includegraphics[width=0.4\textwidth]{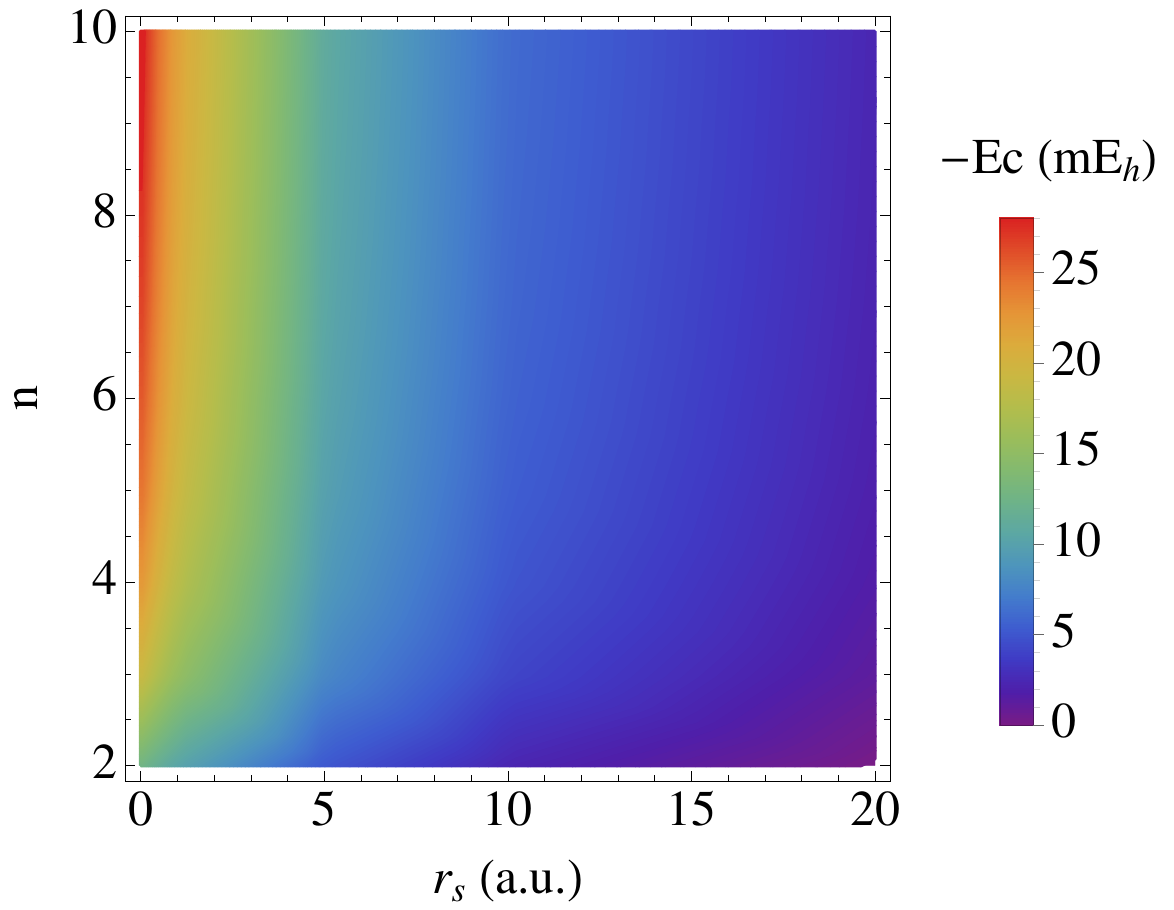}
\caption{
\label{fig:Ec}
\alert{Reduced correlation energies (in $\mEh$) for $n$-ringium with various $r_s$.}}
\end{figure}

Table \ref{tab:Ec} summarizes our best estimates of the correlation energies of $n$-ringium for various $r_s$ \alert{(see also Fig.~\ref{fig:Ec})}.  For $r_s = 0$, we use the exact $\eps_2(n)$ values from Table \ref{tab:eps}.  For $r_s = 0.1$, we use the Pad\'e approximant
\begin{equation}
	\Ec(r_s,n) \approx \frac{\eps_2(n)}{1-[\eps_3(n)/\eps_2(n)] r_s},
\end{equation}
which provides microhartree accuracy. For $n=2$ and $n = 3$, we use the Explicitly Correlated results from Table \ref{tab:rs1}.  For $n = \infty$ and $1\le r_s \le 20$, we use the DMC results from Lee and Drummond. \cite{Lee11a}  For $n=\infty$ and $0.2 \le r_s \le 0.5$, we performed DMC calculations using the CASINO software \cite{CASINO} following the Lee-Drummond methodology.

For the remaining cases ($4 \le n \le 10$ and $r_s \ge 0.2$), we used our own DMC program.  We achieve sub-$\uEh$ uncertainties for $r_s > 1$ (where the electrons become localized and approach a Wigner crystal \cite{Wigner34}) but it is difficult to achieve this for smaller $r_s$, where the uncertainties are 10 -- 40 $\uEh$.

\begin{table}
	\caption{
	\label{tab:Ec}
Reduced correlation energies (in $\mEh$) for $n$-ringium with various $r_s$.  Statistical errors in the last digit of the DMC energies are shown in parentheses.}
	\begin{ruledtabular}
	\begin{tabular}{ccccccccccccccc}
		$n\backslash r_s$
				&		0		&		0.1		&		0.2		&		0.5		&		1		&		5		&		10		&		20		\\
		\hline
		2		&	13.212		&	12.985		&	12.766		&	12.152		&	11.250		&	7.111		&	4.938		&	3.122		\\
		3		&	18.484		&	18.107		&	17.747		&	16.755		&	15.346		&	9.369		&	6.427		&	4.029		\\
		4		&	21.174		&	20.698		&	20.24(2)	&	19.00(1)	&	17.320(1)	&	10.390(0)	&	7.085(0)	&	4.425(0)	\\
		5		&	22.756		&	22.213		&	21.66(2)	&	20.33(1)	&	18.439(1)	&	10.946(0)	&	7.439(0)	&	4.636(0)	\\
		6		&	23.775		&	23.184		&	22.63(2)	&	21.14(1)	&	19.137(1)	&	11.285(0)	&	7.653(0)	&	4.762(0)	\\
		7		&	24.476		&	23.850		&	23.24(2)	&	21.70(1)	&	19.607(1)	&	11.509(0)	&	7.795(0)	&	4.844(0)	\\
		8		&	24.981		&	24.328		&	23.69(3)	&	22.11(1)	&	19.940(1)	&	11.664(0)	&	7.890(0)	&	4.901(0)	\\
		9		&	25.360		&	24.686		&	24.04(2)	&	22.39(1)	&	20.186(1)	&	11.777(0)	&	7.960(0)	&	4.941(0)	\\
		10		&	25.651		&	24.960		&	24.25(4)	&	22.62(1)	&	20.373(1)	&	11.857(0)	&	8.013(0)	&	4.973(0)	\\
		\vdots	&	\vdots		&	\vdots		&	\vdots	&	\vdots	&	\vdots	&	\vdots	&	\vdots	&	\vdots		\\
		$\infty$	&	27.416		&	26.597		&	25.91(1)	&	23.962(1)	&	21.444(0)	&	12.318(0)	&	8.292(0)	&	5.133(0)	\\
	\end{tabular}
	\end{ruledtabular}
\end{table}

\section{Conclusions}
We have studied $n$-ringium using explicitly correlated and quantum Monte Carlo methods.  Using Hylleraas wave functions, we have obtained the near-exact ground-state energy of the $n = 2$ and $n = 3$ systems for various values of the Seitz radius $r_s$.  For $n \ge 4$, we have performed exact-node DMC calculations to find the exact ground-state energies, with statistical errors in the $\uEh$ range.

We have shown that the reduced correlation energy of $n$-ringium is
\begin{equation}
	\Ec(r_s,n) = \eps_2(n) + \eps_3(n) r_ s +  \ldots 
\end{equation}
for high densities, and
\begin{equation}
	\Ec(r_s,n) = \frac{\eta_0(n)-\eps_1(n)}{r_s} + \frac{\eta_1(n)}{r_s^{3/2}} +  \ldots
\end{equation}
for low densities.  Expressions for the coefficients are given in Eqs.~\eqref{eq:MP2-final}, \eqref{eq:eps3n}, \eqref{eq:EW}, \eqref{eps1} and \eqref{eq:eta32}. 

In the thermodynamic limit, we have found that 
\begin{align}
	\Ec(r_s) & = -\frac{\pi^2}{360} + 0.008446\,r_s+  \ldots,
	\\
	\Ec(r_s) & = -\frac{\ln(\sqrt{2\pi})-3/4}{r_s} + \frac{0.359933}{r_s^{3/2}} +  \ldots,
\end{align}
and shown that the ringium and jellium models are equivalent in the thermodynamic limit.
 
This provides a detailed picture of the energy of this new model over a wide range of $n$ and $r_s$ values and we believe that the correlation energies in Table \ref{tab:Ec} are the most accurate yet reported for $n$-ringium.  These systems are distinct uniform electron gases \cite{UEGs12} and can be used to design a new correlation functional for 1D systems.  We will report such a functional in a forthcoming paper. \cite{UEGs3}

\begin{acknowledgements}
The authors thank Neil Drummond and Shiwei Zhang for helpful discussions, the NCI National Facility for a generous grant of supercomputer time. PMWG thanks the Australian Research Council (Grants DP0984806, DP1094170, and DP120104740) for funding. PFL thanks the Australian Research Council for a Discovery Early Career Researcher Award (Grant DE130101441)
\end{acknowledgements}

\section*{Appendix}
For the ground state, the Hamiltonian \eqref{eq:H} can be recast as
\begin{equation} \label{eq:H-rij}
	H = \sum_{i<j}^n \left[ \left(\frac{r_{ij}^2}{4R^2}-1\right) \frac{\partial^2}{\partial r_{ij}^2} + \frac{r_{ij}}{4R^2} \frac{\partial}{\partial r_{ij}} + \frac{1}{r_{ij}} \right]
	+\sum_{\substack{i \neq j\\ i \neq k}}^{n} \sum_{j<k}^{n}\frac{r_{ik}^2 + r_{jk}^2 - r_{ij}^2}{2\,r_{ik}r_{jk}} \frac{r_{ij}^2 + r_{jk}^2 - r_{ik}^2}{2\,r_{ij}r_{jk}} \frac{\partial^2}{\partial r_{ij} \partial r_{ik}}.
\end{equation}
The first term in \eqref{eq:H-rij} contains the two-body parts of the Hamiltonian while the second includes coupling between electron pairs.

The $n$-electron overlap integrals needed in calculations on $n$-ringium can be systematically constructed using the unit-ring Fourier resolution
\begin{equation} \label{eq:rijm}
	r_{ij}^m = \sum_{k=-\infty}^\infty B_{m,k} \ e^{i k \theta_i} e^{- i k \theta_j},
\end{equation}
where
\begin{equation}
	B_{m,k} = \frac{(-1)^k m!}{(m/2+k)!(m/2-k)!}
\end{equation}
is a signed binomial coefficient.  Eq.~\eqref{eq:rijm} is valid for $m \geq 0$ and terminates if $m$ is an even integer.

Resolving each integrand factor, swapping the order of summation and integration, performing the integrations and resumming, often leads to beautiful expressions.  For example, the cyclic $n$-electron integral yields
\begin{equation} \label{eq:cyclic}
	\langle r_{12}^a r_{23}^b r_{34}^c \ldots r_{n,1}^z \rangle = \sum_{k=-\infty}^\infty B_{a,k} B_{b,k} B_{c,k} \ldots B_{z,k},
\end{equation}
which can be written as a $_{n+1}F_n$ hypergeometric function of unit argument. \cite{NISTbook}

In some cases, the sums can be found in closed form, for example,
\begin{gather}
	\langle r_{ij}^a \rangle = B_{a,0},					\\
	\langle r_{ij}^a r_{kl}^b \rangle = B_{a,0} B_{b,0},	\\
	\langle r_{12}^a r_{23}^b r_{31}^c \rangle = \frac{a! \,b! \,c!}{\left(\frac{a+b}{2}\right)! \left(\frac{a+c}{2}\right)! \left(\frac{b+c}{2}\right)!}
													\frac{\left(\frac{a+b+c}{2}\right)!}{\left(\frac{a}{2}\right)! \left(\frac{b}{2}\right)! \left(\frac{c}{2}\right)!} \label{eq:Knuth},
\end{gather}
but this is not possible in general. \cite{BruijnBook}  (Eq.~\eqref{eq:Knuth} uses the result of Problem 62 of Knuth's book. \cite{KnuthBook})  However, sums such as \eqref{eq:cyclic} converge rapidly and are numerically satisfactory.

\end{document}